\documentclass[11pt]{revtex4-1}

\usepackage{graphicx}

\begin{document}

\title{Particle production and universal thermodynamics}

\author{Subhajit Saha\footnote {subhajit1729@gmail.com}}
\author{Subenoy Chakraborty\footnote {schakraborty@math.jdvu.ac.in}}
\affiliation{Department of Mathematics, Jadavpur University, Kolkata 700032, West Bengal, India.}

%%%%%%%%%%%%%%%%%%%%%%%%%%%%%%%%%%%%%%%%%%%%%%%%%%%%%%%%%%%%%%%%%%%%%%%%%%%%%%%%%%%%%%%%%%%%%%%%%%%%%%%%%%%%%%%%%%%%%%%%%%%%
\begin{abstract}
In the present work, particle creation mechanism has been employed to the Universe as a thermodynamical system. The Universe is considered to be a spatially flat FRW model and cosmic fluid is chosen as a perfect fluid with a barotropic equation of state --- $p=(\gamma -1)\rho$. By proper choice of the particle creation rate, expressions for the entropy and temperature have been determined at various stages of evolution of the Universe. Finally, using the deceleration parameter $q$ as a function of the redshift parameter $z$ based on recent observations, the particle creation rate has been evaluated and its variation at different epochs have been shown graphically.

\end{abstract}

\maketitle
%%%%%%%%%%%%%%%%%%%%%%%%%%%%%%%%%%%%%%%%%%%%%%%%%%%%%%%%%%%%%%%%%%%%%%%%%%%%%%%%%%%%%%%%%%%%%%%%%%%%%%%%%%%%%%%%%%%%%%%%%%%%
%\pacs{04.70.Dy, 04.90+e}
%%%%%%%%%%%%%%%%%%%%%%%%%%%%%%%%%%%%%%%%%%%%%%%%%%%%%%%%%%%%%%%%%%%%%%%%%%%%%%%%%%%%%%%%%%%%%%%%%%%%%%%%%%%%%%%%%%%%%%%%%%%%
\section{INTRODUCTION}

The particle creation mechanism was applied in cosmology since 1930's by Schrodinger \cite{Schrodinger1}, to examine its influence on cosmological evolution. Then, after a long gap, Parker \cite{Parker1} and others \cite{Brout1,Brout2,Glansdorff1,Prigogine1,Gao1} have investigated the influence of particle creation process on the structure of cosmological spacetime. In fact, two early universe problems namely initial singularity and huge entropy production can be well addressed by the particle production process. The irreversible nature of particle creation process \cite{Prigogine1} may avoid initial singularity considering an instability of the vacuum while irreversible energy flow from the gravitational field to the created particles may explain the huge entropy flow in the early phase. So, it is reasonable to speculate that nonequilibrium thermodynamical processes may play a crucial role to describe the early universe. 

On the other hand, the common way of explaining the present accelerating phase is the introduction of dark energy (DE). However, the nature of DE is far from being understood and also the major drawback of most of the DE models in the literature have no physical basis and/or many free parameters.

Among other possibilities to explain the present accelerating stage, inclusion of backreaction in the Einstein field equations through an (negative) effective pressure is much relevant in the context of cosmology and the gravitational production of particles [radiation or cold dark matter (CDM)] provides a mechanism for cosmic acceleration \cite{Prigogine1,Calvao1,Lima1,Lima2}. In particular, in comparison with DE models, the particle creation scenario has a strong physical basis-nonequilibrium thermodynamics. Also, the particle creation mechanism not only unifies the dark sectors (DE+DM) \cite{Lima2} but also it contains only one free parameter as we need only a single dark component (DM). Further, statistical Bayesian analysis with one free parameter should be preferred along the hierarchy of cosmological models \cite{Guimaraes1}. So the present particle creation model which simultaneously fits the observational data and alleviates the coincidence and fine-tuning problems, is better compared to the known (one parameter) models, namely, i)~the concordance $\Lambda$CDM which however suffers from the coincidence and fine-tuning problems \cite{Zlatev1,delCampo1,Steinhardt1} and ii) the brane world cosmology \cite{Deffayet1} which does not fit the SNeIa+BAO+CMB (shift parameter) data \cite{Basilakos1}.

Eckart \cite{Eckart1} and Landau and Lifschitz \cite{Landau1} were the pioneers in the formulation of nonequilibrium thermodynamics and are considered as first order deviations from equilibrium. These theories (first order) suffer from serious drawbacks concerning causality and stability due to truncation in first order. Subsequently, Muller \cite{Muller1}, Israel \cite{Israel1}, Israel and Stewart \cite{Israel2}, Pavon et al. \cite{Pavon1} and Hiscock \cite{Hiscock1} developed second order thermodynamical theories. In these theories, the dissipative phenomena like bulk and shear viscosity and heat flux which describe the deviations of a relativistic fluid from local equilibrium, become dynamical variables having causal evolution equations. Also, the evolution equations restrict the propagation speeds of thermal and viscous perturbations to subluminal level and one may get back to the first order theory if second order effects are eliminated (i.e., relaxation time approaches zero).

In the backgroud of the homogeneous and isotropic models of the Universe, bulk viscosity is the only dissipative phenomenon and it may occur either due to coupling of different components of the cosmic substratum \cite{Weinberg1,Straumann1,Schweizer1,Udey1} or due to particle number non-conserving interactions. Here, quantum particle production out of the gravitational field \cite{Zel'dovich1,Murphy1,Hu1} is considered and so the present work is related to the second choice for the occurence of bulk viscosity. Further, we shall concentrate on isentropic (adiabatic) particle production \cite{Prigogine1,Calvao2}, i.e., production of perfect fluid particles whose entropy per particle is constant. However, due to increase of perfect fluid particles, the phase space of the system is enlarged and as a result, there is entropy production. Also, the condition for isentropic process gives a simple relation between the bulk viscous pressure and the particle creation rate and hence the particle production rate is no longer a parameter, it becomes a dynamical variable \cite{Zimdahl2,Zimdahl3}.

In the present work, we start with entropy flow vector due to Israel and Stewart and combine with cosmological particle production characterized by isentropic process. Assuming phenomenologically the particle production rate as a function of the Hubble parameter, cosmological solutions and thermodynamical parameters like entropy and temperature are evaluated at various stages of the evolution (inflationary phase, matter dominated era and late time acceleration) of the Universe. Finally, assuming the deceleration parameter as a function of the redshift variable ($z$), the particle production rate and the thermodynamical parameters are expressed as a function of $z$ and their variations are presented graphically. The paper is organized as follows. Section 2 deals with the second order formulation of Israel and Stewart, the cosmological solutions and thermodynamical variables are presented in section 3 for different stages of evolution of the Universe. Section 4 is devoted to the determination of the particle creation rate and the thermodynamical parameters, both analytically and graphically as a function of the redshift variable $z$. The summary of the work is presented in Section 5.

\section{PARTICLE PRODUCTION IN COSMOLOGY: NONEQUILIBRIUM THERMODYNAMICS}

For a closed thermodynamical system having $N$ particles, the first law of thermodynamics states the conservation of internal energy $E$ as
\begin{equation}
dE=dQ-pdV,
\end{equation}
where $p$ is the thermodynamic pressure, $V$ is any comoving volume and $dQ$ represents the heat received by the system in time $dt$. Now defining $\rho =\frac{E}{V}$ as the energy density, $n=\frac{N}{V}$, the particle number density and $dq=\frac{dQ}{N}$, the heat per unit particle, the above conservation law takes the form
\begin{equation}
d\left(\frac{\rho}{n}\right)=dq-pd\left(\frac{1}{n}\right).
\end{equation} 
Note that this equation (known as Gibb's equation) is also true when the thermodynamical system is not closed, i.e., $N$ is not constant [$N=N(t)$].

We shall now consider an open thermodynamical system where the number of fluid particles is not preserved. So the particle conservation equation
\begin{center}
$N_{;\mu}^{\mu}=0,~~~~~~~\text{i.e.,} \dot{n}+\theta n=0$
\end{center}
is now modified as 
\begin{equation}
\dot{n}+\theta n=n\Gamma,
\end{equation}
where $N^{\mu}=nu^{\mu}$ is the particle flow vector, $u^{\mu}$ is the particle four velocity, $\theta =u_{;\mu}^{\mu}$ is the fluid expansion, $\dot{n}=n_{,\mu}u^{\mu}$ and $\Gamma$ stands for the rate of change of the particle number in a comoving volume $V$, $\Gamma >0$ indicates particle creation while $\Gamma <0$ means particle annihilation. Any non-zero $\Gamma$ will effectively behave as a bulk viscous pressure of the thermodynamical fluid and nonequilibrium thermodynamics should come into the picture.

Thus, in the context of particle creation, we shall consider spatially flat FRW model of the Universe with line element
\begin{equation}
ds^2=-dt^2+a^2(t)[dr^2+r^2d{\Omega _2}^2]
\end{equation}
as an open thermodynamical system. For the cosmic fluid having energy-momentum tensor
\begin{equation}
T_{\mu \nu}=(\rho +p+\Pi)u_\mu u_\nu +(p+\Pi)g_{\mu \nu},
\end{equation}
the Einstein field equations are
\begin{equation}
\kappa \rho =3H^2~~~~~~~~\text{and}~~~~~~~~\kappa (\rho +p+\Pi)=-2\dot{H}
\end{equation}
and the energy conservation relation ${T^{\mu \nu}}_{;\nu}=0$ gives
\begin{equation}
\dot{\rho}+\theta (\rho +p+\Pi)=0.
\end{equation}
Here $\kappa =8\pi G$ is the Einstein's gravitational constant and the pressure term $\Pi$ is related to some dissipative phenomenon (bulk viscosity). However, in the context of particle creation, the cosmic fluid may be considered as a perfect fluid and the dissipative term $\Pi$ as the effective bulk viscous pressure due to particle production. In other words, the cosmic substratum is not a conventional dissipative fluid, rather a perfect fluid with varying particle number. This statement can clearly be understood for the simple case of adiabatic (i.e., isentropic) particle production as follows \cite{Zimdahl3,Chakraborty1}---

Starting from the Gibb's relation [i.e., from Eq. (2)]
\begin{equation}
Tds=d\left(\frac{\rho}{n}\right)+pd\left(\frac{1}{n}\right)
\end{equation}
and using the conservation equations (3) and (7), the entropy variation can be written as 
\begin{equation}
nT\dot{s}=-\Pi \theta -\Gamma (\rho +p),
\end{equation}
where $s$ is the entropy per particle and $T$ is the temperature of the fluid. Now for isentropic process, the entropy per particle remains constant (variable in a dissipative process), i.e., $\dot{s}=0$ and hence we have
\begin{equation}
\Pi =-\frac{\Gamma}{\theta}(\rho +p).
\end{equation}
Thus, the bulk viscous pressure is entirely determined by the particle production rate. So we may say that at least for the adiabatic process, a dissipative fluid is equivalent to a perfect fluid with varying particle number. Further, although $\dot{s}=0$ still there is entropy production due to the enlargement of the phase space of the system (i.e., due to expansion of the Universe in the present context). So the effective bulk pressure does not characterize a conventional nonequilibrium, rather a state having equilibrium properties as well (but not the equilibrium era with $\Gamma =0$).

Now using the Einstein field equations (6), the condition for adiabatic process [i.e., Eq. (10)] can be stated as \cite{Zimdahl3,Chakraborty1}
\begin{equation}
\frac{\Gamma}{\theta}=1+\frac{2}{3\gamma}\left(\frac{\dot{H}}{H^2}\right),
\end{equation}
where $p=(\gamma -1)\rho$ is the equation of state (EoS) for the cosmic fluid.

In an open thermodynamical system, the entropy change $dS$ can be decomposed into an entropy flow $d_f S$ and the entropy creation $d_c S$, i.e., \cite{Hark1}
\begin{equation}
dS=d_f S+d_c S
\end{equation}
with $d_c S \geq 0$. As in a homogeneous system $d_f S=0$ but there is entropy production due to matter creation, so we have
\begin{equation}
\frac{dS}{dt}=\frac{d_c S}{dt}=\frac{d}{dt}(nsV)=S\Gamma ,
\end{equation}
i.e., on integration,
\begin{equation}
S(t)=S_0 \text{exp}\left[3\int _{a_0}^{a}\frac{\Gamma}{\theta} \frac{da}{a}\right],
\end{equation}
with $S_0 =S(t_0)$, $a=a(t_0)$. Further, from Euler's relation,
\begin{equation}
nTs=\rho +p
\end{equation}
and using the conservation relations (3) and (7), the tempertaure of the cosmic fluid is given by
\begin{eqnarray}
T &=& T_0 a^{-3(\gamma -1)} \text{exp}\left[3(\gamma -1)\int _{a_0}^{a}\frac{\Gamma}{\theta}\frac{da}{a}\right] \nonumber \\
&=& T_1 \left(\frac{S}{a^3}\right)^{\gamma -1}, 
\end{eqnarray}
with $T_1=\frac{T_0}{{S_0}^{\gamma -1}}$.

\section{COSMOLOGICAL SOLUTIONS, BEHAVIOUR OF THE THERMODYNAMICAL PARAMETERS AND A FIELD THEORETIC ANALYSIS}

This section is divided into two subsections. In the first subsection, cosmological solutions for phenomenological values of the particle creation rate (as a function of the Hubble parameter) have been determined and thermodynamical parameters (namely, entropy and temperature) have been evaluated. In the next subsection, a field theoretic description has been shown for the above choices of the particle creation rate parameter.

\subsection*{Subsection A}

{\bf Case I:} \textsc{Early Epochs} --- In the very early universe, (starting from a regular vacuum) most of the particle creation effectively takes place and from thermodynamical point of view we have \cite{Lima3}---
\begin{enumerate}
\item At the beginning of expansion, there should be maximal entropy production rate (i.e., maximal particle creation rate) so that the Universe evolves from nonequilibrium thermodynamical state to equilibrium era with the expansion of the Universe.
\item A regular (true) vacuum for radiation initially, i.e., $\rho \rightarrow 0$ as $a \rightarrow 0$.
\item $\Gamma >H$ in the very early universe so that the created radiation behaves as thermalized heat bath and subsequently the creation rate should fall slower than expansion rate and particle creation becomes dynamically insignificant.
\end{enumerate}
Now, according to Gunzig et al. \cite{Gunzig1}, the simplest choice satisfying the above requirements is that the particle creation rate is proportional to the energy density \cite{Shapiro1}, i.e., $\Gamma =3\beta \frac{H^2}{H_r}$, where $H_r$ has the dimension of Hubble parameter (i.e., reciprocal of time). Thus $\beta$ has the dimension of $M^{-1}T^{-1}$.

For this choice of $\Gamma$, the Hubble parameter can be obtained from Eq. (11) as \cite{Chakraborty1}
\begin{equation}
H=\frac{H_r}{\beta +(1-\beta){\left(\frac{a}{a_r}\right)}^{\left(\frac{3\gamma}{2}\right)}},
\end{equation}
where $\beta$ is the constant of proportionality and $H_r$ is the Hubble parameter at some fixed time $t_r$ [with $a_r =a(t_r)$]. Thus, as $a \rightarrow 0$, $H\rightarrow {\beta}^{-1}H_r=\text{constant}$, indicating an exponential expansion ($\ddot{a}>0$) in the inflationary era while for $a \gg a_r$, $H \propto a^{-\frac{3\gamma}{2}}$ represents the standard FRW cosmology ($\ddot{a}<0$). Suppose $a_r$ is identified as some intermediate value of $a$ where $\ddot{a}=0$ (i.e., the transition epoch from de Sitter stage to standard radiation era). Then we have $\dot{H}_r=-{H_r}^2$ and from Eq. (11)
\begin{equation}
\beta =1-\frac{2}{3\gamma}.
\end{equation}
Hence for relativistic matter (i.e., $\gamma =\frac{4}{3}$ and $\beta =\frac{1}{2}$),
\begin{equation}
H=\frac{2H_r}{1+\left(\frac{a}{a_r}\right)^2},
\end{equation}
which on integration gives \cite{Zimdahl1}
\begin{equation}
t=t_r+\frac{1}{4H_r}\left[\text{ln} \left(\frac{a}{a_r}\right)^2 +\left(\frac{a}{a_r}\right)^2 -1\right].
\end{equation}
So in the limiting situation \cite{Schweizer1},
\begin{eqnarray}
a &\simeq& {a_r} e^{2H_r t}~, \text{for}~~ a \ll a_r ~(\text{Inflationary era}), \nonumber \\
a &\simeq& {a_r} t^{\frac{1}{2}}~,\text{for}~~ a \gg a_r ~(\text{Standard cosmological regime}). \nonumber
\end{eqnarray}
Thus, in the standard cosmological regime, the rate of particle production $\Gamma$ decreases as inverse square law, i.e., $\Gamma \sim t^{-2}$. Also from Eqs. (14) and (16), the expressions for the thermodynamical parameters are 
\begin{equation}
S(t)={S_r} \left[(1-\beta)+\beta \left(\frac{a}{a_r}\right)^{-\frac{3\gamma}{2}}\right]^{-\frac{2}{\gamma}},
\end{equation}
and
\begin{equation}
T(t)=T_r \left[\beta +(1-\beta) \left(\frac{a}{a_r}\right)^{\frac{3\gamma}{2}}\right]^{-\frac{2(\gamma -1)}{\gamma}}
\end{equation}
respectively, where $S_r=S(t_r)$ and $T_r=T(t_r)$. Eqs. (21) and (22) imply
\begin{eqnarray}
S &\sim& \left(\frac{a}{a_r}\right)^3,~~~T \sim \text{const.},~~\text{for} ~a \gg a_r \nonumber \\
S &\sim& \text{const.},~~~~~T \sim \left(\frac{a}{a_r}\right)^{-3(\gamma -1)},~~\text{for} ~a \ll a_r. \nonumber
\end{eqnarray}

Further, integrating (7) and using (10), the energy density has the expression
\begin{equation}
\rho =\rho _r \left[\beta +(1-\beta){\left(\frac{a}{a_r}\right)}^{\frac{3\gamma}{2}}\right]^{-2},
\end{equation}
i.e.,~~ $T=T_0 \rho ^{\frac{\gamma -1}{\gamma}}$, ~$T_0 =\frac{T_r}{{\rho _r}^{\frac{\gamma -1}{\gamma}}}$. 

Thus we see that in the radiation era (i.e., $\gamma =\frac{4}{3}$), we have $\rho \propto T^4$, i.e.,
the Universe as a thermodynamical system behaves as a black body \cite{Lima3,Modak1}.\\\\
{\bf Case II:} \textsc{Intermediate Decelerating Phase} --- Here, the simplest natural choice is $\Gamma \propto H$ \cite{Carneiro1}. It should be noted that this choice of $\Gamma$ does not satisfy the third thermodynamical requirement at the early universe (mentioned above). Also, the solution [see Eq. (24)] will not satisfy the above condition (ii).

For this choice of $\Gamma$, one can integrate Eq. (11) to obtain
\begin{equation}
H^{-1}=\frac{3\gamma}{2}(1-\Gamma _0)t,~~~a={a_0} t^{l},~l=\frac{2}{3\gamma (1-\Gamma _0)},
\end{equation}
which is the usual power law expansion of the Universe in standard cosmology with particle production rate decreases as $t^{-1}$. Hence $H$ (i.e., $\rho$) does not satisfy the true vacuum condition. Also, the entropy grows as power law
\begin{equation}
S=S_0 \left(\frac{t}{t_0}\right)^{l\Gamma _0}
\end{equation}
while the expression for temperature is given by
\begin{equation}
T=T_0 a^{(\Gamma _0-3)(\gamma -1)}.
\end{equation}
This temperature may be thought of as analogous to the reheating temperature in standard models. Also note that the expression for the temperature becomes constant in dust era (i.e., $\gamma =1$) or when $\Gamma _0=3$. For $\Gamma _0>3$, the temperature gradually increases with the evolution of the Universe till dust era and then it gradually decreases.\\\\
{\bf Case III:} \textsc{Late Time Accelerated Expansion} --- In this case, the thermodynamical requirements of Case I are modified as---
\begin{enumerate}
\item There should be minimum entropy production rate at the beginning of the late time accelerated expansion and the Universe again becomes thermodynamically nonequilibrium.
\item The late time false vacuum should have $\rho \rightarrow 0$ as $a \rightarrow \infty$.
\item The creation rate should be faster than the expansion rate.
\end{enumerate}
We shall now show that another simple choice of $\Gamma$, namely $\Gamma \propto \frac{1}{H}$ will satisfy these requirements. This choice of $\Gamma$ gives the Hubble parameter [by integrating Eq. (11)] as \cite{Chakraborty1}
\begin{equation}
H^2=\delta H_f +\left(\frac{a}{a_f}\right)^{-3\gamma},
\end{equation}
where $\delta$ is the constant of proportionality and $a_f$ is the value of the scale factor at the instant when the Universe enters the quintessence era. Integrating again, we have from (27),
\begin{equation}
a=a_f \left[\frac{1}{\sqrt{\delta H_f}} \text{sinh} \left\lbrace \frac{3\gamma}{2} \sqrt{\delta H_f}(t-t_f)\right\rbrace \right]^{\left(\frac{2}{3\gamma}\right)}.
\end{equation}
Hence for $a \ll a_f$, $H \sim a^{-\frac{3\gamma}{2}}$, i.e., the usual power law expansion in standard cosmology while $H \sim \sqrt{\delta H_f}$ when $a \gg a_f$, i.e., the accelerated expansion at late time (in quintessence era). So in the matter dominated era (i.e., $a \ll a_f$), $\Gamma$ grows as $a^{\frac{3\gamma}{2}}$ and gradually it approaches the constant value $(\delta H_f)^{-\frac{1}{2}}$ in the quitessence era. For this choice of $\Gamma$, the thermodynamical parameters are given by
\begin{equation}
S=S_f \left[\text{cosh} \left\lbrace \frac{3\gamma}{2} \sqrt{\delta H_f}(t-t_f)\right\rbrace \right]^{\left(\frac{2}{\gamma}\right)}
\end{equation}
and
\begin{equation}
T=\frac{T_f}{\sqrt{\delta H_f}} \left[\text{coth} \left\lbrace \frac{3\gamma}{2} \sqrt{\delta H_f}(t-t_f)\right\rbrace \right]^{\frac{2(\gamma -1)}{\gamma}}.
\end{equation}
Note that at the instant $t=t_f$, i.e., the beginning of the quintessence era, the entropy function has the minimum value while the temperature blows up there. Hence the first modified thermodynamical requirement is satisfied. Also the above solution shows that $\rho \rightarrow 0$ as $a \rightarrow \infty$. Therefore, this choice of $\Gamma$ correctly describes the late time acceleration.

\subsection*{Subsection B}

%\begin{small} \begin{center} {\bf Scalar Field Model: Field Theoretic Analysis} \end{center} \end{small}

If the present particle creation model is equivalent to a minimally coupled scalar field $\phi$ having potential $V(\phi)$, then Eq. (11) modifies to
\begin{equation}
\frac{\Gamma}{\theta}=1-\frac{w(\phi)}{\gamma},
\end{equation}
where $w(\phi)=1+\frac{p(\phi)}{\rho (\phi)}$ is the effective EoS parameter for the scalar field and $\rho (\phi)=\frac{1}{2}\dot{\phi}^2 +V(\phi)$ and $p(\phi)=\frac{1}{2}\dot{\phi}^2 -V(\phi)$ are the energy density and thermodynamic pressure for the scalar field. Note that the effective EoS parameter $w(\phi)$ is constant when $\Gamma \propto H$ while for the other two cases (i.e., $\Gamma \propto H^2$ or $\Gamma \propto \frac{1}{H}$), the effective EoS is variable.  Further, the ratio of the state parameters will characterize whether there will be creation or annihilation of particles. In particular, if the kinetic energy (K.E.) of the scalar field dominates over its potential energy (P.E.), i.e., $w(\phi) \sim 2$, then there is always particle annihilation when $\gamma <2$. On the other hand, if the P.E. of the scalar field dominates over its K.E. (scalar field behaves as a cosmological constant), i.e., $w(\phi) \sim 0$, then there will always be particle creation with $\Gamma =3H$. However, if $p_{\phi}=0$, i.e., $w(\phi) =1$, then particle creation or annihilation will depend on whether $\gamma >1$ or $\gamma <1$.
%In particular, if the K.E. of the scalar field dominates over its potential energy and $\gamma <1$, then there is always particle annihilation while if the potential energy of the scalar field dominates over its K.E. term (i.e., $p_{\phi}$ is negative) and $\gamma >1$, then always there will be particle creation. Finally, the scalar field behaves as a cosmological constant when $\Gamma =3H$.

\section{PARAMETRIC CHOICE OF DECELERATION PARAMETER AND ESTIMATION OF PARTICLE CREATION RATE AND THERMODYNAMIC PARAMETERS}

From Eq. (11), using the definition of $q=-(1+\frac{\dot{H}}{H^2})$, one can express the particle creation rate $\Gamma$ in terms of the deceleration parameters as
\begin{equation}
\Gamma =3H\left[1-\frac{2}{3\gamma}(1+q)\right],
\end{equation}
where the expression for the Hubble parameter is
\begin{equation}
H=H_0 \text{exp}\left[\int _{0}^{z} \frac{1+q}{1+z}dz\right]
\end{equation}
with $z=\frac{1}{a}-1$ as the redshift parameter. Also, the entropy and temperature of the cosmic fluid can be expressed in terms of the deceleration parameter as
\begin{equation}
S=S_0 \text{exp}\left[3\int _z^{z_0}\left\lbrace 1-\frac{2}{3\gamma}(1+q)\right\rbrace \frac{dz}{(1+z)}\right]
\end{equation}
and
\begin{equation}
T=T_1 \lbrace{S(1+z)^3\rbrace}^{\gamma -1}.
\end{equation}
It is worthwhile to study the variation of $\Gamma$ for some parametric approximation of the deceleration parameter along the cosmic evolution. We shall use the following three parametrization of the deceleration parameter.
\begin{enumerate}
\item $q(z)={q_0}+{q_1}z$,
\item $q(z)={q_0}+{q_1}\frac{z}{1+z}$, 
\item $q(z)=\frac{1}{2}+\frac{{q_0}z+{q_1}}{(1+z)^2}$,
\end{enumerate}
where the parameters are estimated from observations.

\subsection{$q(z)={q_0}+{q_1}z$}

This is the simplest two parameter linear parametrization of the deceleration parameter given by Riess et al. \cite{Riess1} and Cunha \cite{Cunha1}. Here $q_0$ is the present value of the deceleration parameter and ${q_1}=\frac{dq}{dz}|_{z=0}$ and their best fit values are $-0.73$ and $1.5$ respectively. Then the explicit expression of $H$, $\Gamma$ and the thermodynamical parameters are
\begin{eqnarray}
H(z) &=& {H_0}e^{{q_1}z} (1+z)^{1+q_0 -q_1}, \\
\Gamma (z) &=& 3H_0 e^{q_1 z} (1+z)^{1+q_0 -q_1} \left[1-\frac{2}{3\gamma}(1+{q_0}+{q_1}z)\right], \\
S(z) &=& S_0 \text{exp}\left[3\left\lbrace 1-\frac{2}{3\gamma}(1+q_0)\right\rbrace \text{ln} \left(\frac{1+z_0}{1+z}\right)-\frac{2q_1}{\gamma}\left\lbrace (z_0-z)-\text{ln}\left(\frac{1+z}{1+z_0}\right)\right\rbrace \right], \\
T(z) &=& T_1 {\left\lbrace S_0 (1+z)^3 \text{exp}\left[3\left\lbrace 1-\frac{2}{3\gamma}(1+q_0)\right\rbrace \text{ln} \left(\frac{1+z_0}{1+z}\right)-\frac{2q_1}{\gamma}\left\lbrace (z_0-z)-\text{ln}\left(\frac{1+z}{1+z_0}\right)\right\rbrace \right] \right\rbrace}^{\gamma -1}. 
\end{eqnarray}
Figures (1a)--(1d) show the graphical representation of $H$, $\Gamma$, and the thermodynamical parameters $S$ and $T$ for this choice of $q(z)$. In drawing the figures, various choices of $\gamma$ have been considered which correspond to different phases, namely, $\gamma =\frac{4}{3}$ for radiation, $\gamma =\frac{1}{3}$ seems to represent a fluid of wall domains, $\gamma =1$ represents dust, $\gamma =-\frac{1}{3}$ represents a phantom fluid. From the figures, it can be seen that the phantom case leads to negative values of $\Gamma$ in the future, i.e., it leads to particle annihilation.

\begin{figure}
\begin{minipage}{0.4\textwidth}
\includegraphics[width=1.0\linewidth]{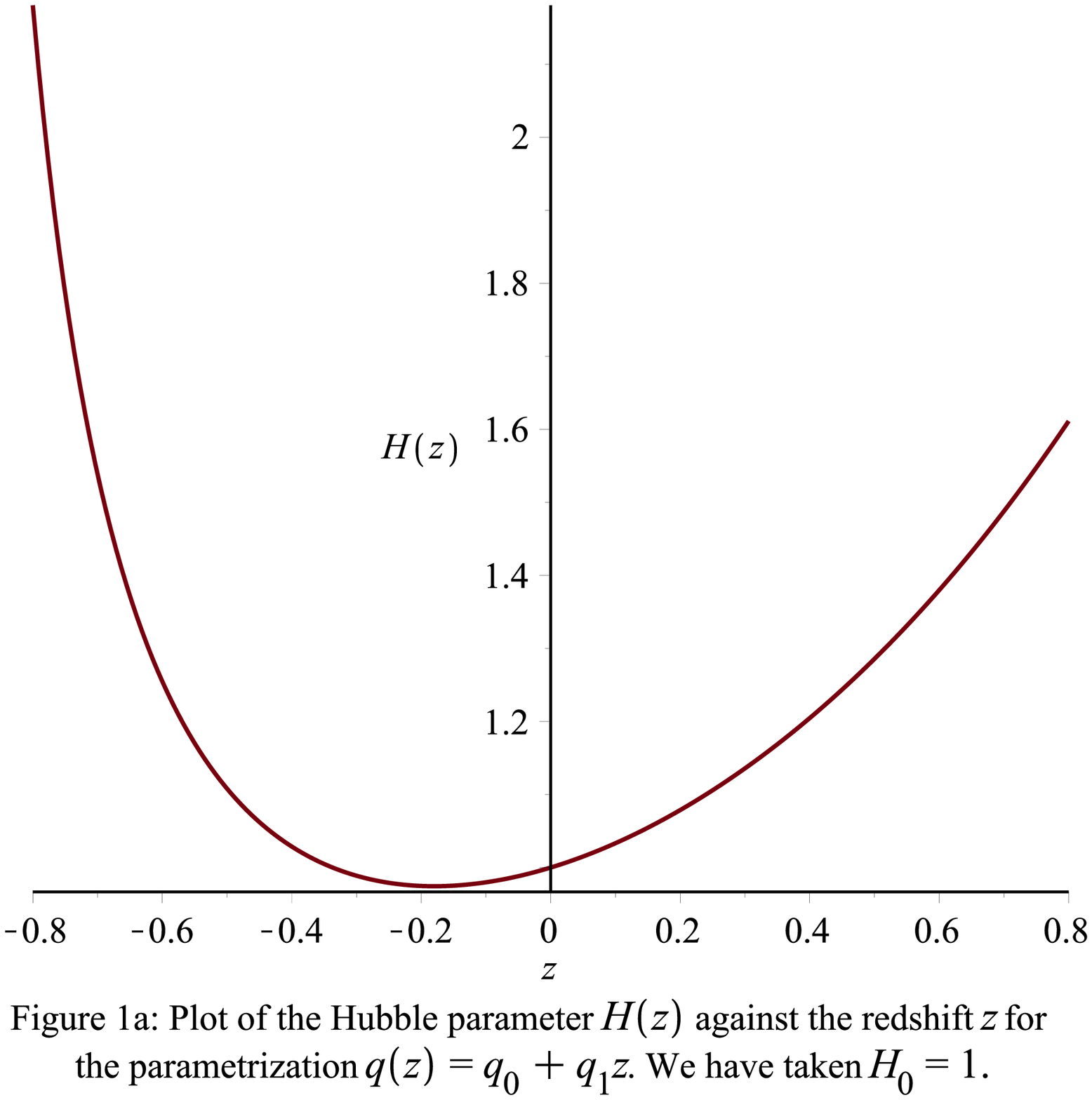}
\end{minipage}
\begin{minipage}{0.4\textwidth}
\includegraphics[width=1.0\linewidth]{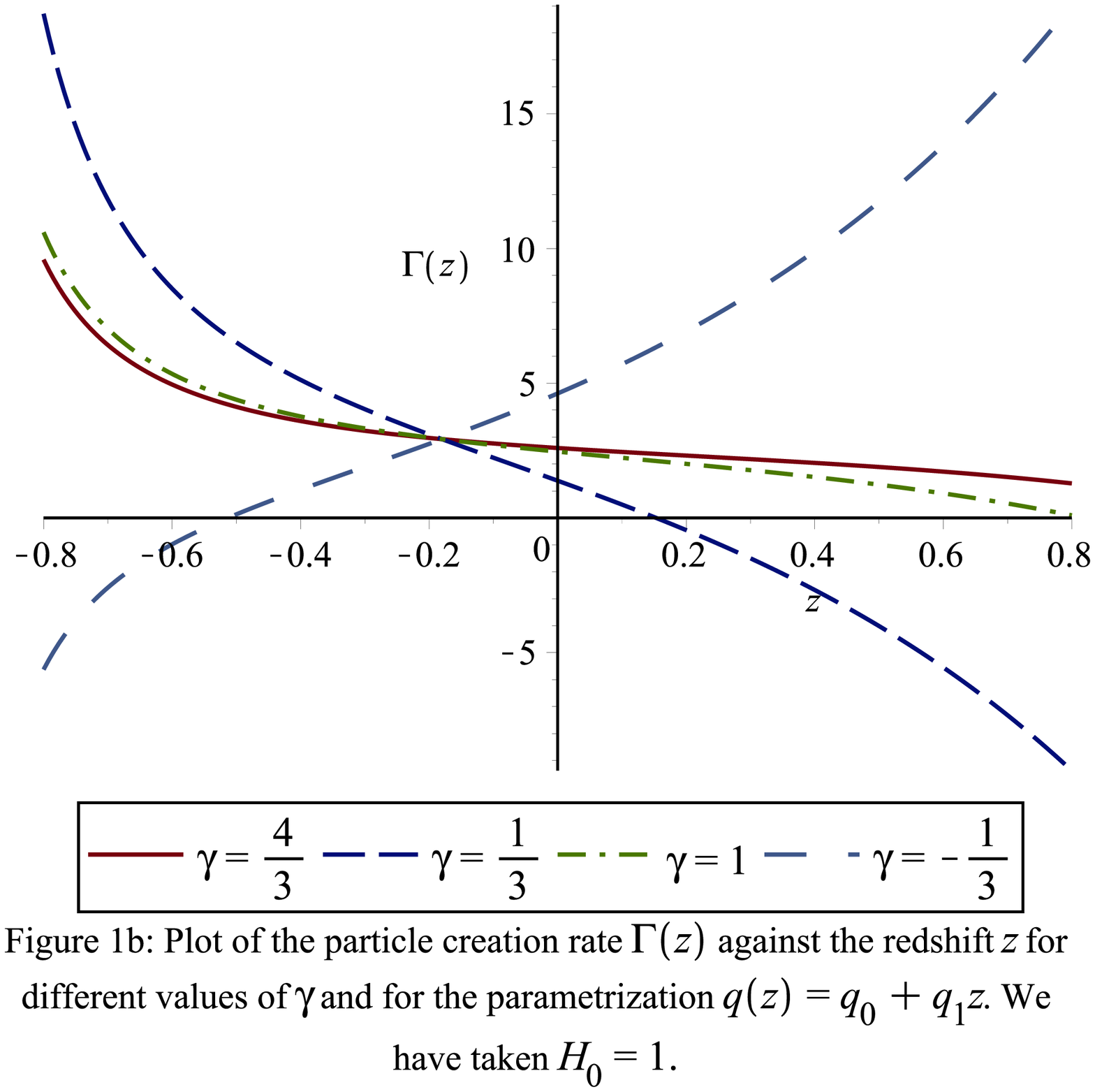}
\end{minipage}
\end{figure}

\begin{figure}
\begin{minipage}{0.4\textwidth}
\includegraphics[width=1.0\linewidth]{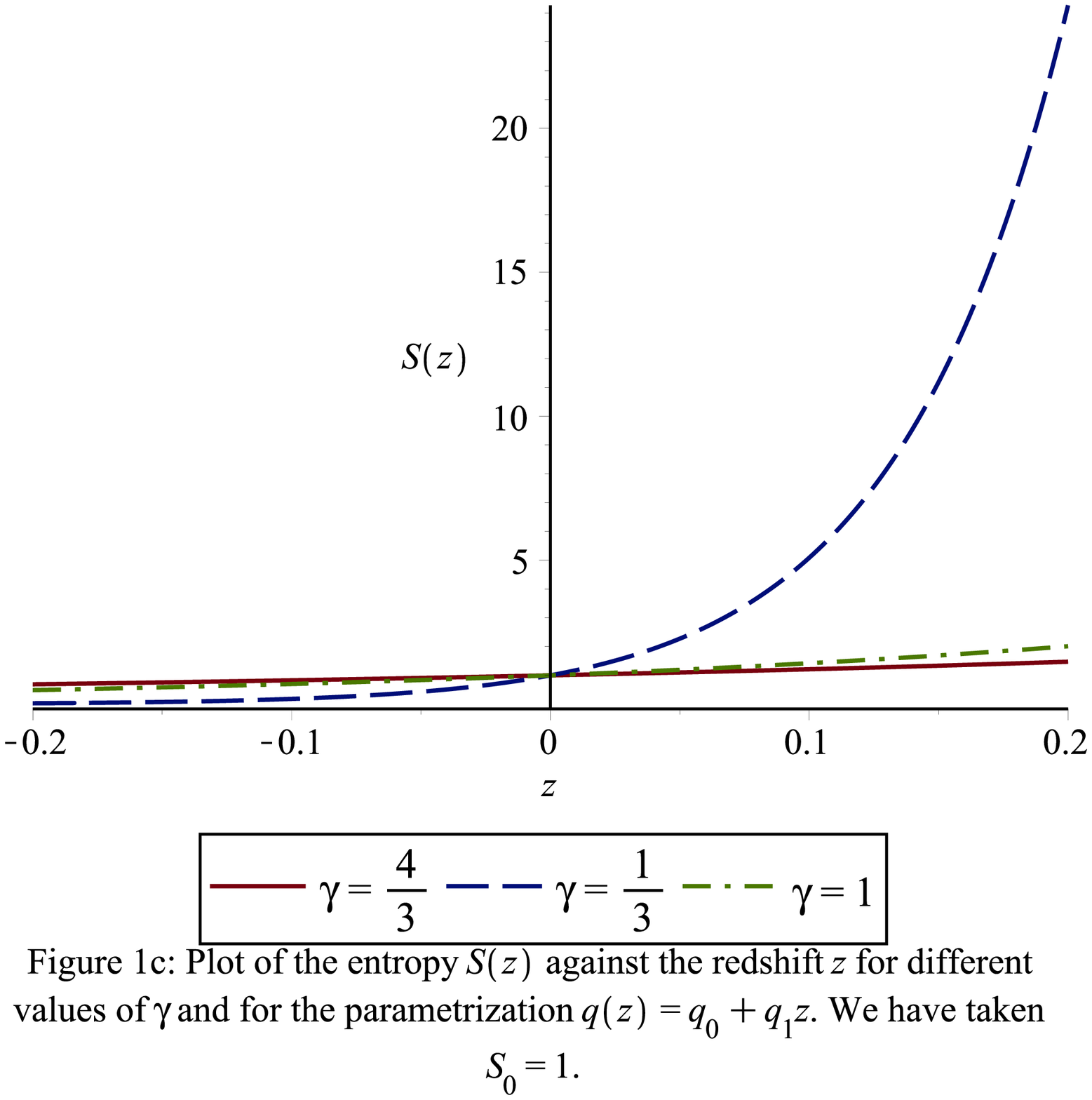}
\end{minipage}
\begin{minipage}{0.4\textwidth}
\includegraphics[width=1.0\linewidth]{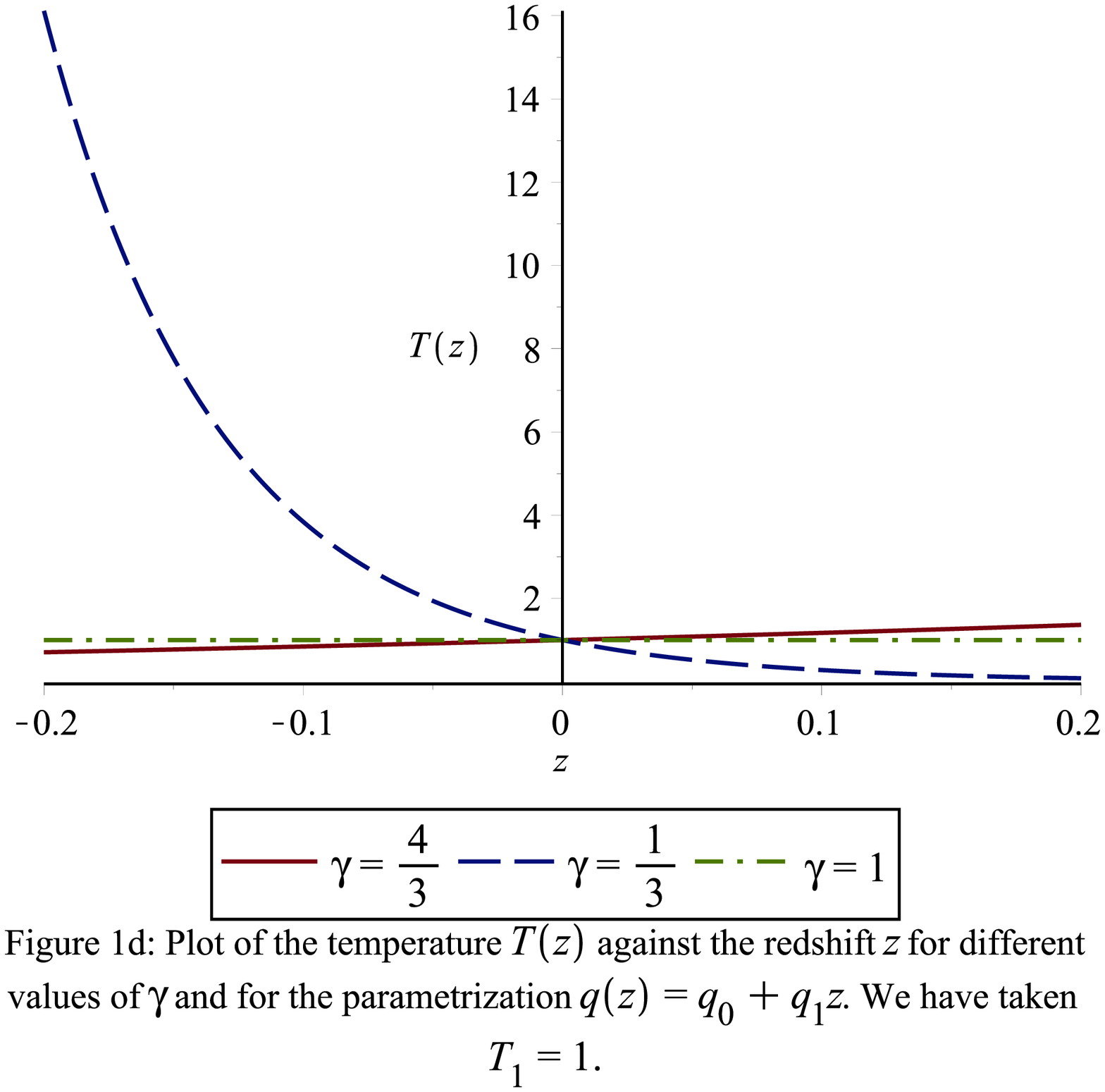}
\end{minipage}
\end{figure}

\subsection{$q(z)={q_0}+{q_1}\frac{z}{1+z}$}

This parametric representation of $q(z)$ is based on the recent SNeIa observational data \cite{Mortsell1,Sollerman1,Santos1}, namely, Union 2 sample of 557 events \cite{Amanullah1,Alcaniz1,Costa1,Jesus1,Costa2}. For the present flat model, the estimated values of ($q_0,q_1$) are ($-0.66$ $\pm$ $0.33$($1\sigma$)$\pm$ $0.07$($2\sigma$), $1.54$ $\pm$ $0.19$($1\sigma$)$\pm$ $0.38$($2\sigma$)). Then $H$, $\Gamma$, $S$ and $T$ can be written as a function of $z$ as
\begin{eqnarray}
H(z) &=& {H_0}e^{-q_1 \frac{z}{1+z}}(1+z)^{1+q_0 +q_1}, \\
\Gamma (z) &=& 3H_0 e^{-q_1 \frac{z}{1+z}}(1+z)^{1+q_0 +q_1} \left\lbrace 1-\frac{2}{3\gamma}\left(1+q_0 +q_1 \frac{z}{1+z}\right)\right\rbrace , \\
S(z) &=& S_0 \text{exp}\left[3\left\lbrace 1-\frac{2}{3\gamma}(1+q_0)\right\rbrace \text{ln}\left(\frac{1+z_0}{1+z}\right)-\frac{2q_1}{\gamma}\left\lbrace \text{ln}\left(\frac{1+z_0}{1+z}\right)+\frac{1}{1+z_0}-\frac{1}{1+z}\right\rbrace \right], \\
T(z) &=& T_1 {\left\lbrace S_0 (1+z)^3 \text{exp}\left[3\left\lbrace 1-\frac{2}{3\gamma}(1+q_0)\right\rbrace \text{ln}\left(\frac{1+z_0}{1+z}\right)-\frac{2q_1}{\gamma}\left\lbrace \text{ln}\left(\frac{1+z_0}{1+z}\right)+\frac{1}{1+z_0}-\frac{1}{1+z}\right\rbrace \right]\right\rbrace}^{\gamma -1}.
\end{eqnarray}
The graphical representation of the above parameters have been shown in figures (2a)--(2d) respectively.

\begin{figure}
\begin{minipage}{0.4\textwidth}
\includegraphics[width=1.0\linewidth]{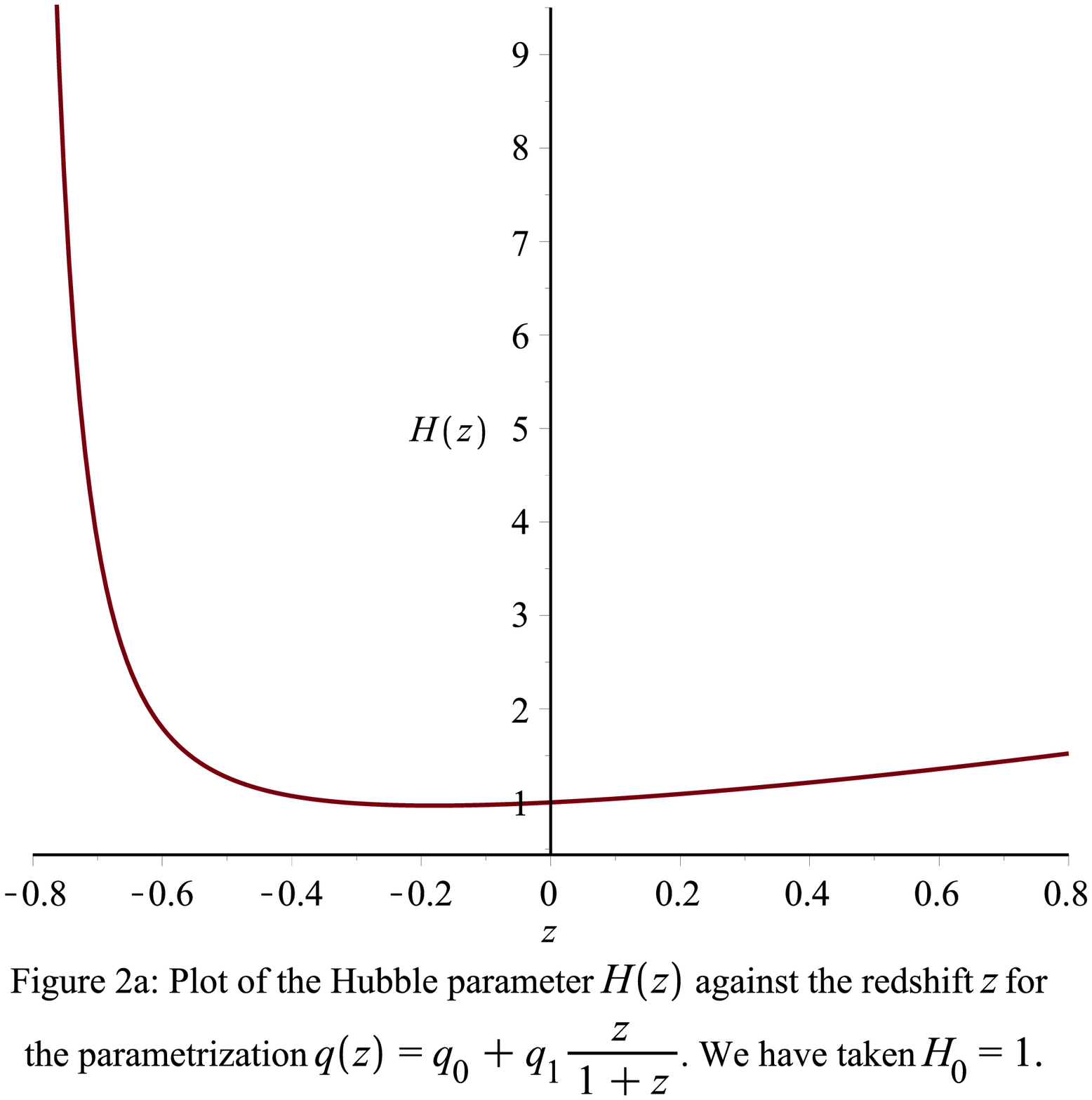}
\end{minipage}
\begin{minipage}{0.4\textwidth}
\includegraphics[width=1.0\linewidth]{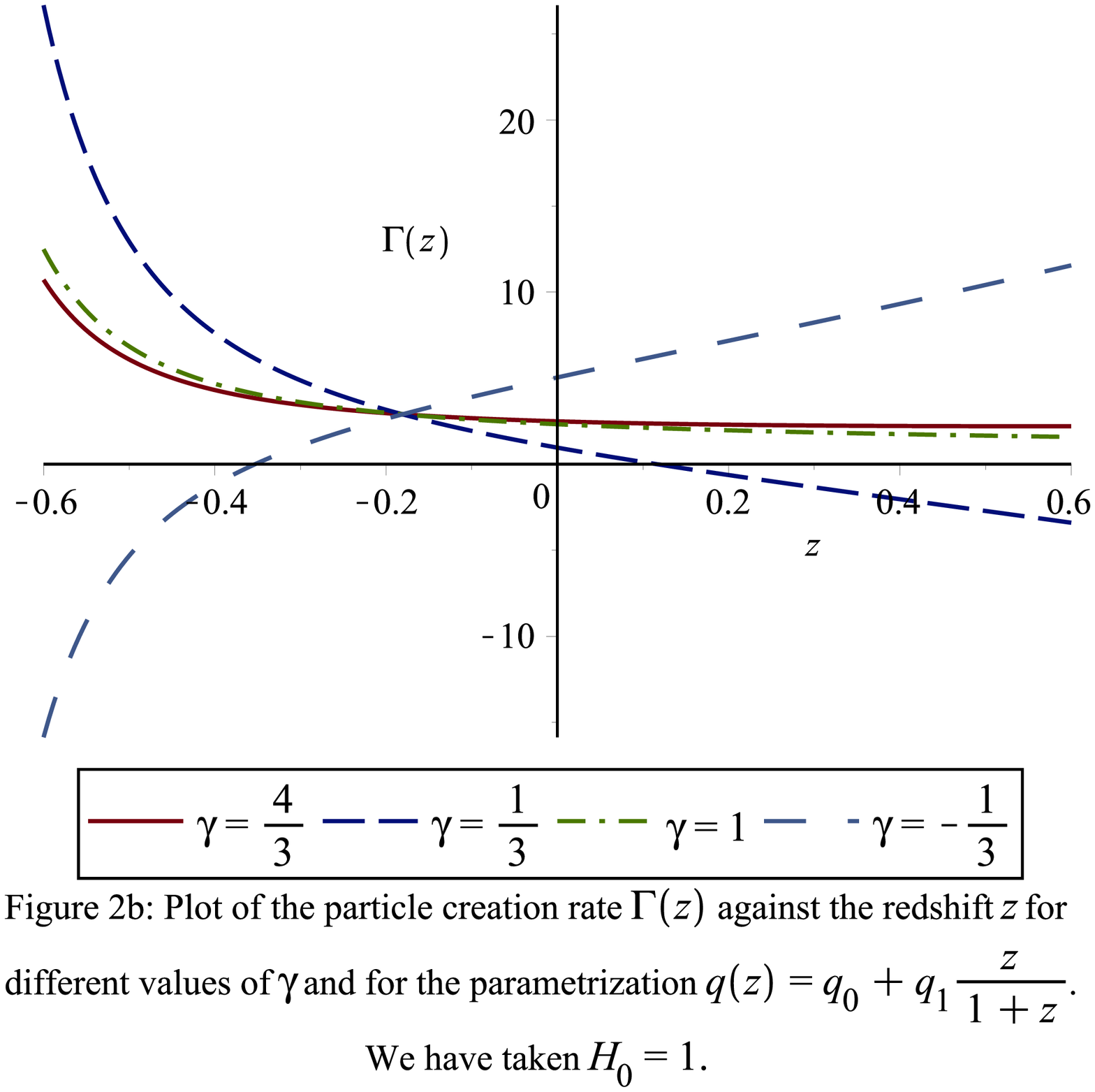}
\end{minipage}
\end{figure}

\begin{figure}
\begin{minipage}{0.4\textwidth}
\includegraphics[width=1.0\linewidth]{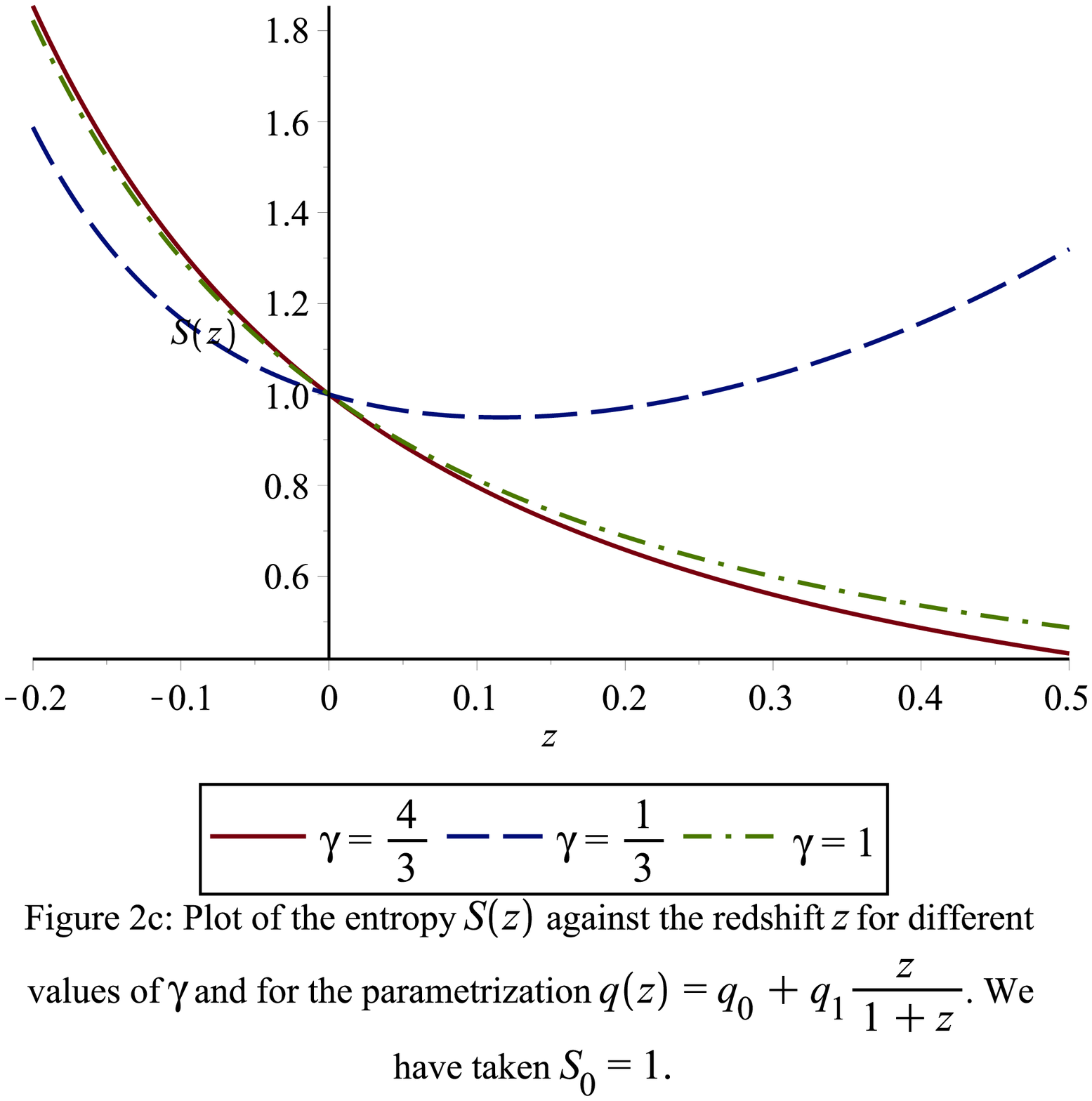}
\end{minipage}
\begin{minipage}{0.4\textwidth}
\includegraphics[width=1.0\linewidth]{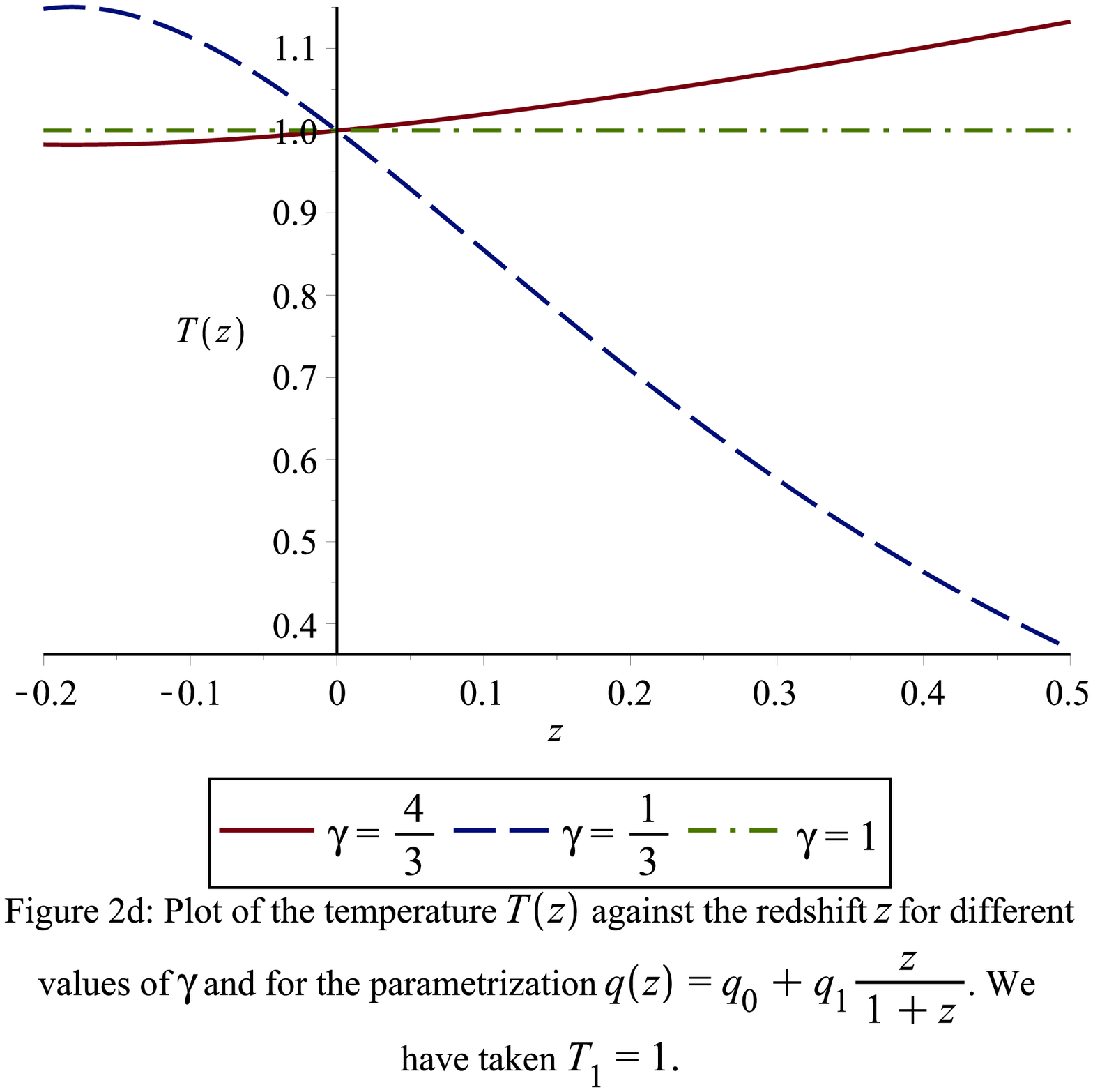}
\end{minipage}
\end{figure}

\subsection{$q(z)=\frac{1}{2}+\frac{{q_0}z+{q_1}}{(1+z)^2}$}

This is another parametrization of $q(z)$ common in the literature \cite{Gong1,Gong2}. Here the present value of the deceleration parameter is ($\frac{1}{2}+q_1$) and the best fit values are $q_0 =1.47$ and $q_1 =-1.46$. Using Eqs. (32)--(35), the explicit form of $H$, $\Gamma$ and the thermodynamical parameters are
\begin{eqnarray}
H(z) &=& H_0 (1+z)^{\frac{3}{2}} \text{exp}\left[\frac{q_1}{2}+\frac{{q_0}z^2 -{q_1}}{2(1+z)^2}\right], \\
\Gamma (z) &=& 3H_0 (1+z)^{\frac{3}{2}} \text{exp}\left[\frac{q_1}{2}+\frac{{q_0}z^2 -{q_1}}{2(1+z)^2}\right]\left[1-\frac{2}{3\gamma}\left\lbrace \frac{3}{2}+\frac{{q_0}z+{q_1}}{(1+z)^2}\right\rbrace \right], \\
S(z) &=& S_0 \text{exp} \Biggl[3\left(1-\frac{1}{\gamma}\right)\text{ln}\left(\frac{1+z_0}{1+z}\right)-\frac{1}{\gamma}(q_0+q_1) \left\lbrace \frac{1}{(1+z_0)^2}-\frac{1}{(1+z)^2}\right\rbrace -\frac{2q_0}{\gamma}\Biggl\{ \frac{1}{1+z} \nonumber \\
&-& \frac{1}{1+z_0}\Biggr\} \Biggr], \\
T(z) &=& T_1 \Biggl\{ S_0 (1+z)^3 \text{exp} \Biggl[3\left(1-\frac{1}{\gamma}\right)\text{ln} \left(\frac{1+z_0}{1+z}\right)-\frac{1}{\gamma}(q_0+q_1) \left\lbrace \frac{1}{(1+z_0)^2}-\frac{1}{(1+z)^2}\right\rbrace  \nonumber \\
&-& \frac{2q_0}{\gamma} \Biggl\{ \frac{1}{1+z}-\frac{1}{1+z_0}\Biggr\} \Biggr]\Biggr\}^{\gamma -1}.
\end{eqnarray}
In figures (3a)--(3d), the variation of $H$, $\Gamma$, $S$, and $T$ have been presented for this particular parametrization of $q(z)$. 

\begin{figure}
\begin{minipage}{0.4\textwidth}
\includegraphics[width=1.0\linewidth]{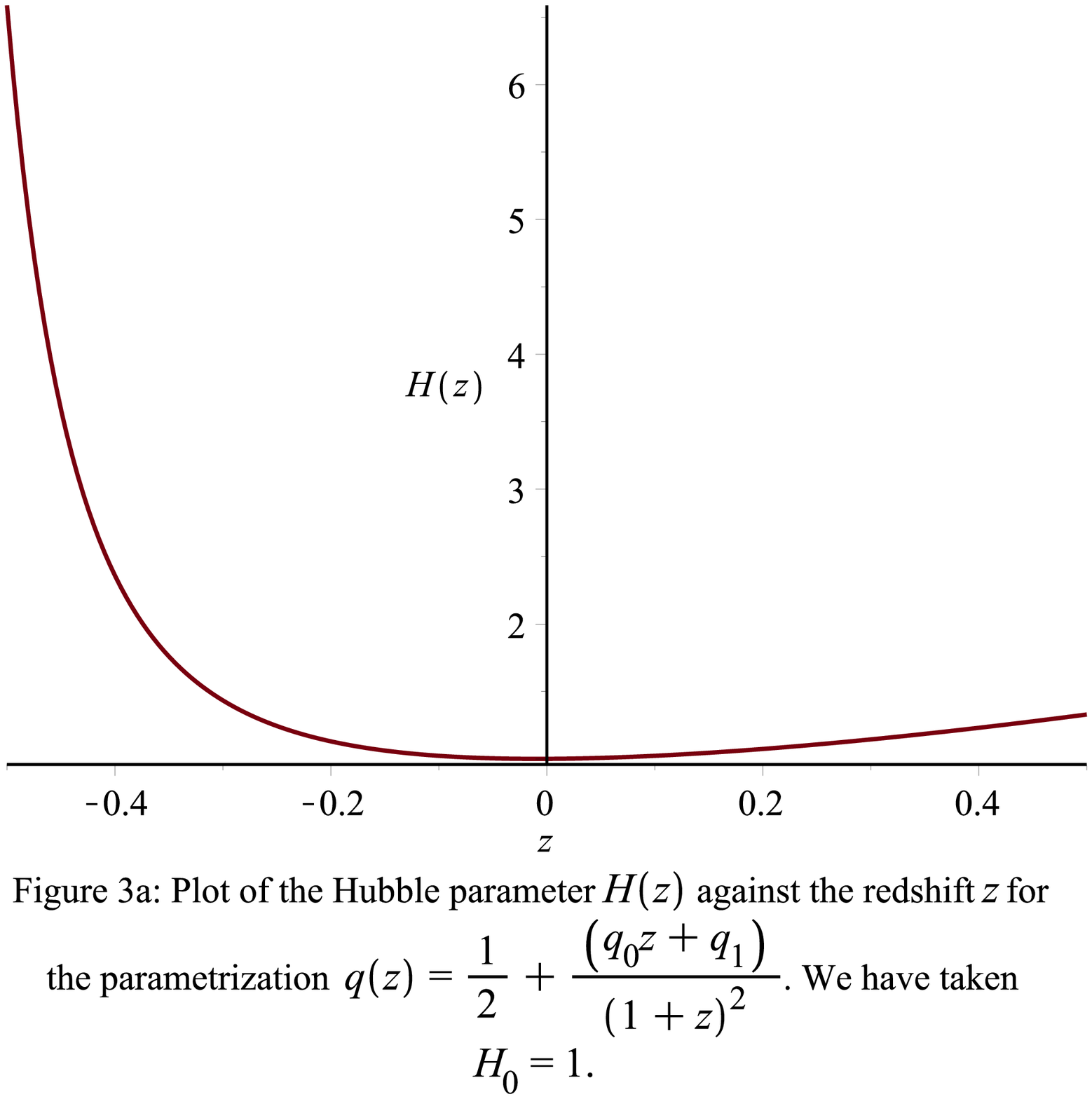}
\end{minipage}
\begin{minipage}{0.4\textwidth}
\includegraphics[width=1.0\linewidth]{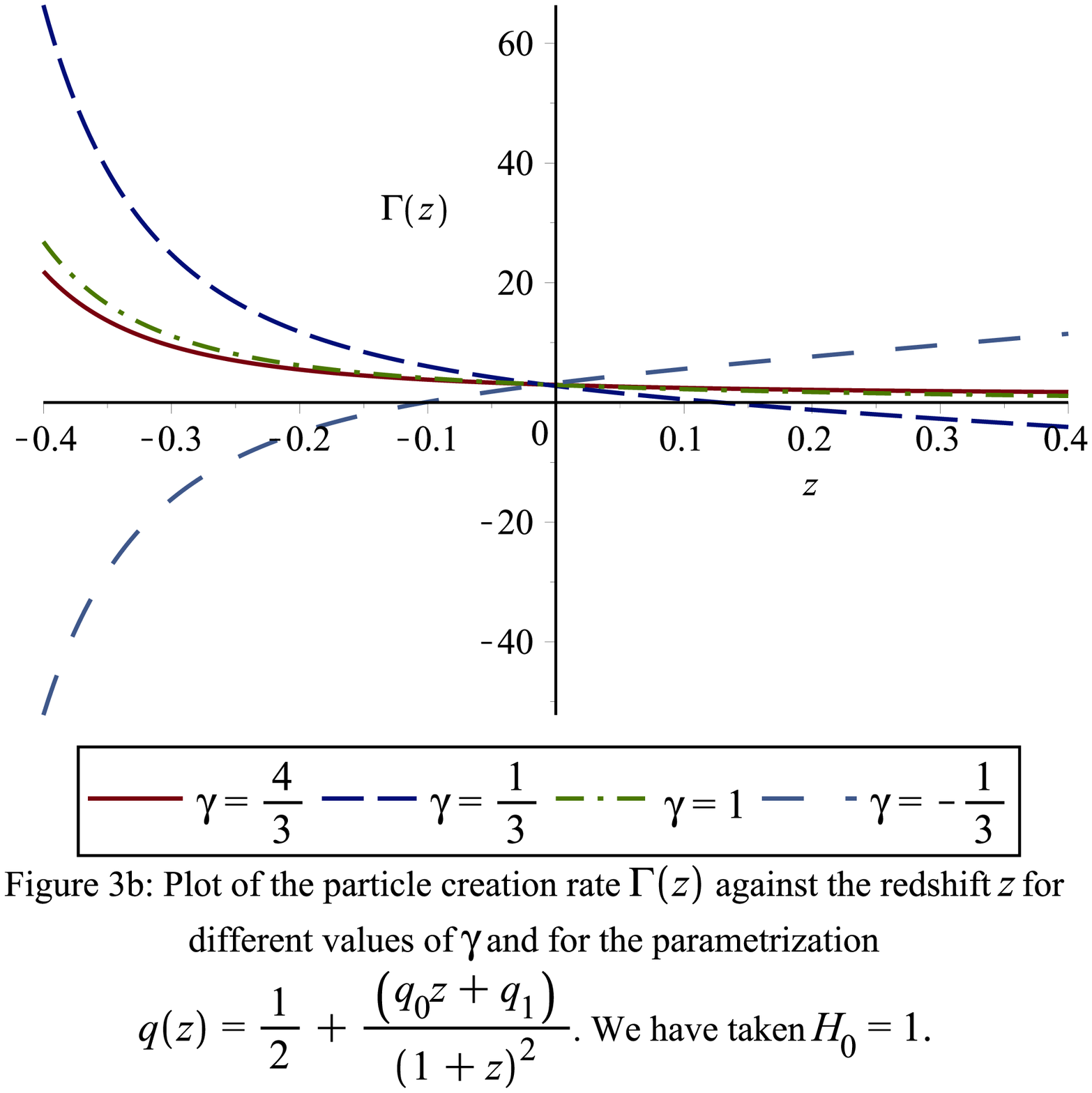}
\end{minipage}
\end{figure}

\begin{figure}
\begin{minipage}{0.4\textwidth}
\includegraphics[width=1.0\linewidth]{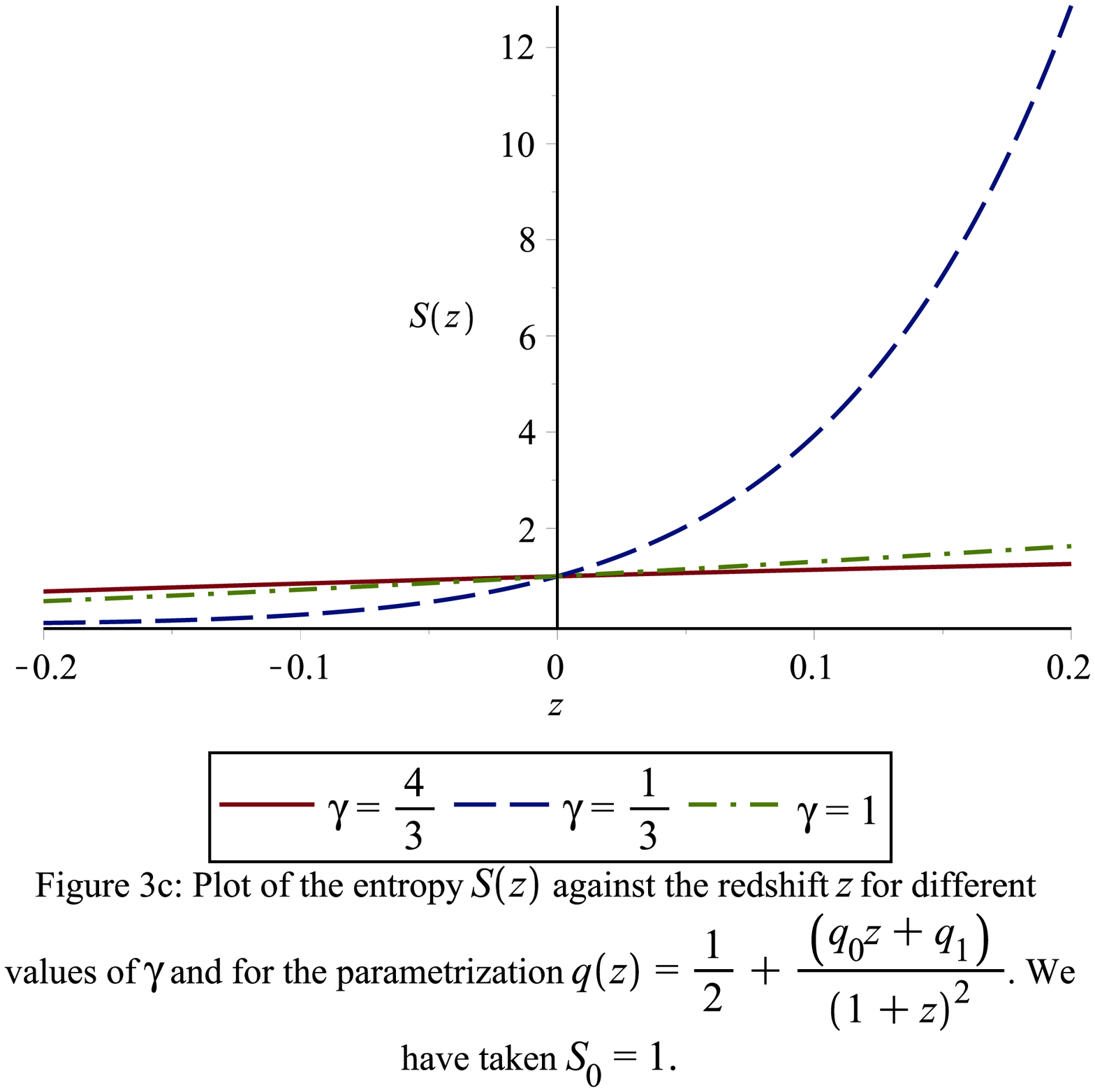}
\end{minipage}
\begin{minipage}{0.4\textwidth}
\includegraphics[width=1.0\linewidth]{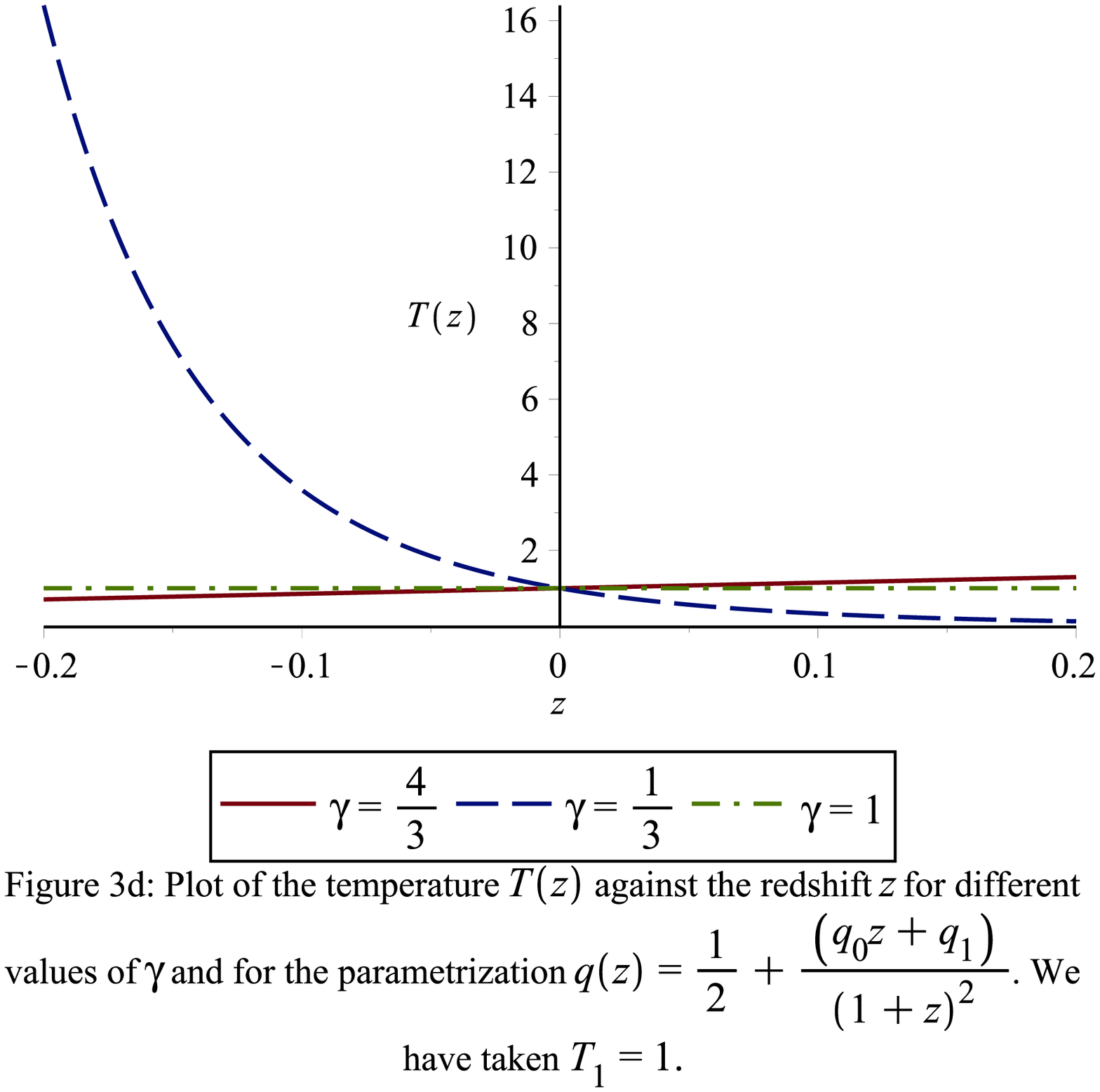}
\end{minipage}
\end{figure}

In the following, we have chosen interacting holographic dark energy (HDE) \cite{Li1,Wang1} as the cosmic fluid. It is a DE model based on the holographic principle. From the effective quantum field theory, one can obtain the holographic energy density. The expression for $\gamma$ in interacting HDE model can be written as
\begin{equation}
\gamma =1-b^2-\frac{\Omega _d}{3}\left(1+\frac{2\sqrt{\Omega _d}}{c}\right),
\end{equation}
where $\Omega _d$ is the density parameter of HDE and $c$ is a dimensionless parameter. Both the quantities can be obtained from the Planck data sets. Also, the parameter $b^2$ is known as the interacting (or coupling) parameter and is related to the interaction term as $Q=3b^2H\rho$. Using three Planck data sets \cite{Li2,Saha1}, we have evaluated the bounds on the interacting (or coupling) parameter $b^2$ for which $\gamma$ becomes negative. The results have been presented in Table I.\\

\begin{center} {\bf Table I}: Bounds on $b^2$ (which make $\gamma$ negative) for different {Planck} data sets \end{center} 
\begin{center}
\begin{tabular}{lccc}
\hline \hline Data & $c$ & ${\Omega}_{d}$ & Bounds on $b^2$\\
\hline Planck+CMB+SNLS3+lensing & 0.603 & 0.699 & $b^2 > 0.1209$\\
	   Planck+CMB+Union 2.1+lensing & 0.645 & 0.679 & $b^2 > 0.1954$\\
	   Planck+CMB+BAO+HST+lensing & 0.495 & 0.745 & All values of $b^2$\\
\hline \hline
\end{tabular}
\end{center}

\begin{figure}
\includegraphics[width=7.0cm,height=7.0cm]{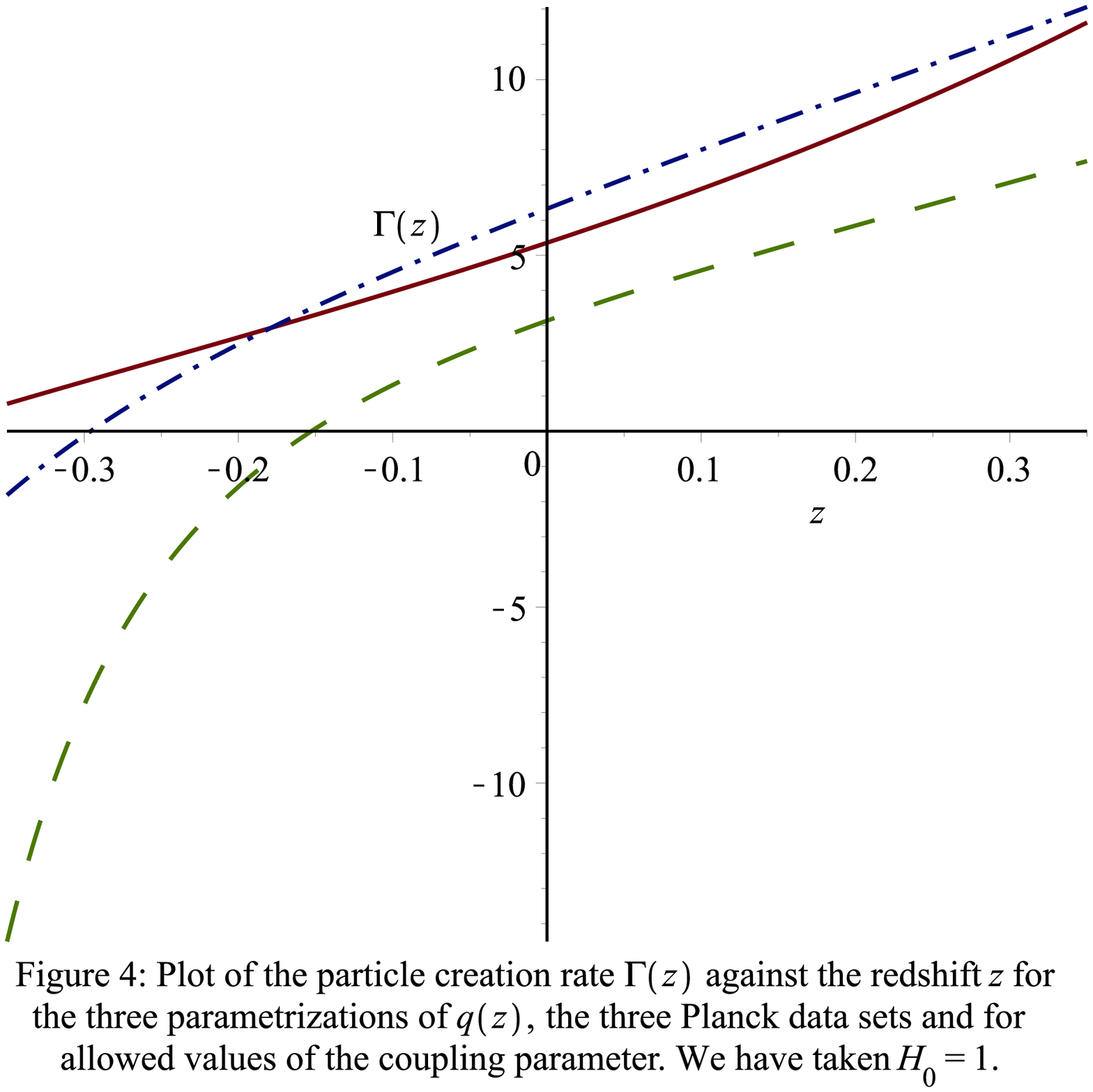}
\end{figure}

\vspace*{0.2cm}

From the above table and Fig. 4, it is evident that for these bounds on $b^2$, the particle production rate $\Gamma$ diminishes in the phantom domain and the thermodynamical equilibrium of the Universe bounded by the event horizon supports these bounds on $b^2$ \cite{Saha1}.

\section{SUMMARY OF THE WORK}

The present work deals with universe as a nonequilibrium thermodynamical system with dissipation due to particle creation process. The Universe is chosen as spatially flat FRW spacetime and the cosmic substratum is chosen as a perfect fluid with a barotropic EoS. Phenomenologically as well as thermodynamically the particle creation rate is estimated by a function of the Hubble parameter and cosmological solutions are evaluated. These solutions correspond to inflationary epoch, matter dominated era and late time acceleration of the Universe. Also thermodynamical parameters, namely entropy and temperature are determined at the above phases of evolution of the Universe. It is found that at the radiation era (i.e., $\gamma =\frac{4}{3}$), the Universe behaves as a black body (i.e., $\rho \propto T^4$) and is in agreement with the references \cite{Lima3,Modak1}. The temperature in the intermediate decelerating phase can be considered as reheating temperature in standard models. The field theoretic correspondence shows restriction on K.E. or P.E. of the analogous scalar field for particle creation (or annihilation) mechanism. Finally, in section 4, we proceed in the reverse way. We start with three parametric choices of the deceleration parameter based on recent observations and the creation rate $\Gamma$, the Hubble parameter $H$ and the thermodynamical parameters $S$ and $T$ are evaluated as a function of the redshift parameter as well as their variations are presented graphically. The thermodynamic parameters entropy and temperature have similar behaviour when the given fluid is chosen as normal fluid (i.e., fluid obeying strong energy condition) while for exotic fluid, parameters have distinct character. From the figures, we see that the entropy decreases with the late time evolution of the Universe for $\gamma =\frac{1}{3}$ while it is almost constant for $\gamma =\frac{4}{3}$ and $\gamma =1$ for the first and third parametric choice of the deceleration parameter while for the second parametric choice of the deceleration parameter, the entropy function increases for $\gamma =\frac{4}{3}$ and $\gamma =1$ and it has a minimum in the recent past and then it increases again for $\gamma =\frac{1}{3}$. On the other hand, the temperature increases for $\gamma =\frac{1}{3}$ for all the three parametric choices of the deceleration parameter, but it is almost constant for $\gamma =\frac{4}{3}$ and $\gamma =1$. Therefore, from the above analysis, it can be said that the Universe again becomes nonequilibrium thermodynamical system in the late time accelerated expansion.

From the Figs. 1(b), 2(b) and 3(b), we see that the particle creation rate increases with the evolution of the Universe. This is not unusual if we carefully look at Eq. (32). As long as the cosmic fluid satisfies the WEC (in the quintessence era), i.e., $\gamma >0$, $\Gamma$ will increase with decrease of $q$ and it is reflected in the graphical representation of $\Gamma$. However, if $\gamma$ is negative, i.e., cosmic fluid violates the WEC (in phantom domain) and then $\Gamma$ decreases with decrease of $q$ as shown in Figs. 1(b), 2(b) and 3(b) with spacedashed line graph for $\gamma =-\frac{1}{3}$. Particularly, choosing the cosmic fluid to be interacting HDE and using observed values of the parameters $c$ and $\Omega _d$ from three Planck data sets, we have obtained bounds on the coupling parameter $b^2$ in table I. For these bounds, $\gamma$ is negative and the particle production rate diminishes in the phantom domain. Lastly, in Ref. \cite{Saha1}, the bounds on $b^2$ for thermodynamical equilibrium of the Universe bounded by event horizon supports the bounds given in Table I. Therefore, we may conclude that as long as the cosmic fluid is in quintessence era, there is an increase of particle production rate but it decreases when the cosmic fluid is considered in phantom domain.  

%%%%%%%%%%%%%%%%%%%%%%%%%%%%%%%%%%%%%%%%%%%%%%%%%%%%%%%%%%%%%%%%%%%%%%%%%%%%%%%%%%%%%%%%%%%%%%%%%%%%%%%%%%%%%%%%%%%
\begin{acknowledgments}

The authors are thankful to IUCAA, Pune, India for their warm hospitality and research facilities as the work has been done there during a visit. Also SC acknowledges the UGC-DRS Programme in the Department of Mathematics, Jadavpur University. The author SS is thankful to UGC-BSR Programme of Jadavpur University for awarding research fellowship.

\end{acknowledgments}
%%%%%%%%%%%%%%%%%%%%%%%%%%%%%%%%%%%%%%%%%%%%%%%%%%%%%%%%%%%%%%%%%%%%%%%%%%%%%%%%%%%%%%%%%%%%%%%%%%%%%%%%%%%%%%%%%%%%%%%%%%%%

\end{document}